# Mismatch error correction for time interleaved analog-to-digital converter over a wide frequency range

Zouyi Jiang,[1,2] Lei Zhao,[1,2,a)] Xingshun Gao[1,2], Ruoshi Dong[1,2], Jinxin Liu[1,2], and Qi An[1,2]

[1]*State Key Laboratory of Particle Detection and Electronics, University of Science and Technology of China, Hefei 230026, China*

[2]*Department of Modern Physics, University of Science and Technology of China, Hefei 230026, China*

High-speed high-resolution Analog-to-Digital Conversion is the key part for waveform digitization in physics experiments and many other domains. This paper presents a new fully digital correction of mismatch errors among the channels in Time Interleaved Analog-to-Digital Converter (TIADC) systems. We focus on correction with wide-band input signal, which means that we can correct the mismatch errors for any frequency point in a broad band with only one set of filter coefficients. Studies were also made to show how to apply the correction algorithm beyond the base band, i.e. other Nyquist zones in the under-sampling situation. Structure of the correction algorithm is presented in this paper, as well as simulation results. To evaluate the correction performance, we actually conducted a series of tests with two TIADC systems. The results indicate that the performance of both two TIADC systems can be greatly improved by correction, and the Effective Number Of Bits (ENOB) is successfully improved to be better than 9.5 bits and 5.5 bits for an input signal up to the bandwidth (-3dB) range in the 1.6-Gsps 14-bit and the 10-Gsps 8-bit TIADC systems, respectively. Tests were also conducted for input signal frequencies in the second Nyquist zone, which shows that the correction algorithms also work well as expected.

## I. INTRODUCTION

High-speed waveform digitization allows access to the most detailed information of interesting signals, and is widely used in many domains, such as nuclear and particle physics experiments[1], data communication[2], measurement instruments[3] and medical imaging[4].

For high-speed waveform digitization, the Time-Interleaved Analog-to-Digital Converter (TIADC) is a well-known technology to achieve higher sample rate based on current ADC technology. However, the mismatches containing offset, gain and sample-time errors among all sub-ADC channels decrease the performance of Spurious Free Dynamic Range (SFDR) and ENOB, and limit the development of TIADC technology. Previous published studies of mismatch correction for TIADC technology have been widely incorporated into ultra-high speed ADC chip design and waveform digitization system.

Analog correction is one of the mismatch correction methods, which can reduce or even eliminate the mismatches.[5-7] It can correct mismatches from the source of mismatch generation. But a high-precision feedback circuit is essential, which is complicated and difficult to design. Besides, it's also usually difficult to achieve the purpose of completely eliminating the mismatches for the limitation of feedback adjustment accuracy.

Digital correction is the other mismatch correction method, which is focused on the effects of mismatch errors and eliminate or reduce the influence in the digital domain. Although the mismatch errors still exist, its influence has been weakened or even eliminated after the digital correction.

Lots of digital correction methods have been presented in the early researches and articles, of which, two are corresponding to the mismatch calibration methods. One is applied in foreground calibration, suitable for the application with stable temperature and supply voltage. In these stable conditions, the mismatches are stable and can be calibrated before operating. After calibrating the mismatch errors, many correction methods are applicable in the foreground calibration, such as the interpolation[8], fractional delay filter[9], Generation Wavelets[10], perfect reconstruction methods[11] and so on. With Look-Up-Tables (LUTs), the correction elements can calibrate the temperature dependence and design the correction parameters before normal operation, which can extend the application conditions.

The other one is used in background calibration.[12-20] The mismatches are not needed to be known before normal work. The correction element not only corrects the mismatch errors but also measures the mismatches in time. The self-adaptive method is a best choice in this situation.

a) Author to whom correspondence should be addressed: zlei@ustc.edu.cn

In order to measure the mismatches without interrupting the normal operation, Ref. 12 uses another sub-ADC as a reference. In Ref. 13, it needs a special processing for the ADC input circuits and the spectrum component of input signal is limited in Refs. 14 and 15.

Although the self-adaptive method can be applied in a complex operating condition, it uses too much hardware resource to implement on Field Programmable Gate Arrays (FPGAs), Digital Signal Processors (DSPs) or even Application Specific Integrated Circuits (ASICs) for real-time correction with a high speed. And it is hard to follow the variation immediately when the input signal shape changes too fast.

For the under-sampling situation, there are few articles discussing it. In Ref. 21, it gives a hardware correction implementation structure, but it is hard to apply at a very high sampling rate without a parallel structure. While in Ref. 22, it uses much more resource to implement the feedback structure.

The perfect reconstruction method is useful in the narrowband correction. In Ref.23, a method was proposed to address the issue for wide band correction, but no hardware implementation and verification was conducted, and it is only suitable for the input signal within the baseband. Besides, it uses both analog and digital filter banks, which makes its structure much more complex and the analog part very sensitive to environmental variation and noise. There are no articles jet discussing the perfect reconstruction applying in under-sampling. In this paper, we introduce a new fully digital method based on perfect reconstruction theory for broadband mismatch correction not only in normal sampling but also in under-sampling mode. And we realize the mismatch correction of the 1.6-Gsps 14-bit TIADC system[24] and 10-Gsps 8-bit TIADC system.

## II. DIGITAL CORRECTION

### A. Mismatches in TIADC System

In a TIADC system, there are M sub-ADCs. Each sub-ADC is sampling by a clock with frequency of $F_s/M$, where $F_s$ is the sampling frequency of the TIADC system. And the phases of M sampling clocks are evenly distributed in the periodic of $F_s/M$ with a $2\pi/M$ phase shift. Then we can get the samples of the M ADCs in order, which is equivalent to get the samples by a signal ADC with sampling frequency of $F_s$. For an instance, in a TIADC system with four sub-ADCs as shown in Fig. 1, the phase shift between adjacent sub-ADCs is 90 degrees, in the other words, the clock phases are 0, 90, 180 and 270 degrees.

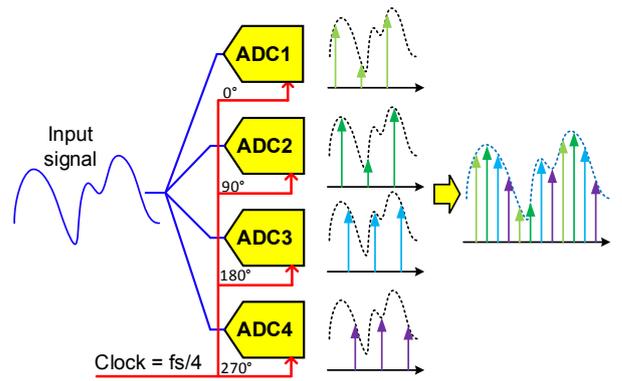

Fig. 1. The architecture of TIADC system

In an ideal condition, the input signal is distributed to the M sub-ADCs with the same amplitude and phase. But in the real system, it is almost impossible to achieve. The gain and delay mismatches among the M channels both exist. $H_m(j\Omega)$ is used to express the response of the real distribution circuits. And we also consider the time shift of sampling clock in $H_m$ to make all the sub-ADCs sampling with the same clock to simplify the analysis. So the response of the analog circuits is presented as (1),

$$H_m(j\Omega) = g_m e^{j\Omega t_m} = g_m e^{j\Omega(mT_S + \Delta t_m)} \quad (1)$$

where $g_m$ is the gain of the $m^{th}$ (m = 0,1, …, M-1) channel consisting of the gain of ADC and the gain of the distributed circuits before ADC. $\Delta t_m$ is the time error with reference to the ideal sample time. In this paper, $g_m$ and $\Delta t_m$ are both frequency dependent as shown in (2), where $\Omega$ is the frequency of input signal. The $T_s$ is the sample clock period of the TIADC system. And $e^{j\Omega mT_s}$ is represented the phase shift of sampling-clock shown in Fig.1.

$$g_m = g_m(\Omega), \Delta t_m = \Delta t_m(\Omega) \quad (2)$$

After the $m^{th}$ ADC, $x_m[n]$ the digital output of this channel is given as (3)

$$x_m[n] = u_m(t_n) = g_m u_m(n \times T_1 + m \times T_S + \Delta t_m) + \Delta o_m \quad (3)$$

where $T_1$ is the real sampling clock period, which is the multiply of $T_s$ by M. And $\Delta o_m$ is the offset of $m^{th}$ channel, which is usual constant referring to the frequency of input signal and we do not discuss it in detail in the later part as it is easy to be corrected. The direct digital output of the TIADC system is y[n], which is consisted of the digital outputs of M channel ($x_m[n]$) in order. With the summation format, we can use an M up-sampling element and a delay element to get the ordered output samples as shown in Fig. 2, which is the schematic diagram of TIADC system consisting of both the analog-to-digital conversion and the recombination of digital output.

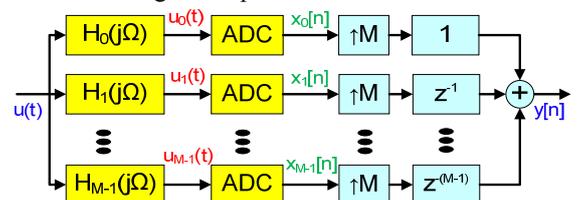

Fig. 2. The schematic diagram of TIADC principle

It is difficult to directly analyze the error of y[n], so we solve the problem in frequency domain later. Based on Fig.2 the DFT of y[n] is given as (4) when ignoring the offset error. In (4), ω is the digital frequency of the input signal and ω=ΩT$_s$.

$$Y(e^{j\omega}) = DFT(y[n]) = \sum_{m=0}^{M-1} e^{-j\omega m} X_m(e^{jM\omega})$$

$$= \frac{1}{MT_S} \sum_{k}^{\infty} (U(j(\frac{\omega}{T_S} - \frac{2\pi k}{MT_S})) \times \sum_{m=0}^{M-1} (e^{-j\omega m} H_m(j(\frac{\omega}{T_S} - \frac{2\pi k}{MT_S})))) \quad (4)$$

$$= \frac{1}{MT_S} \sum_{k}^{\infty} (U(j(\frac{\omega}{T_S} - \frac{2\pi k}{MT_S})) \times \sum_{m=0}^{M-1} (g_m e^{j(\omega - 2\pi k/M)(\Delta t_m/T_S)} e^{-j2\pi km/M}))$$

In an ideal condition (i.e. $g_m = 1$, $\Delta t_m = 0$), (4) is equal to the Nyquist-Shannon Sampling Theorem with a sampling clock period of $T_S$, which is almost impossible to be achieved.

Based on the analysis above, the key of the correction is how to eliminate the pseudo spectrum from the digital output. However, both the gain and sampling-time errors are contributed to the pseudo spectrum and we can't separate them. And it is complex to be expressed because of the frequency dependence of mismatches, which makes the research more challenging.

### B. Correction Method

#### 1) The basic structure of correction

Referring to the research and contribution of the perfect reconstruction method[14], we use M filters to correct the gain and sampling-time errors and reconstruct the output. The filter response is given as $F_m(e^{j\omega})$ and the correction structure is shown as Fig. 3, where the delay elements are considering in the correction filters.

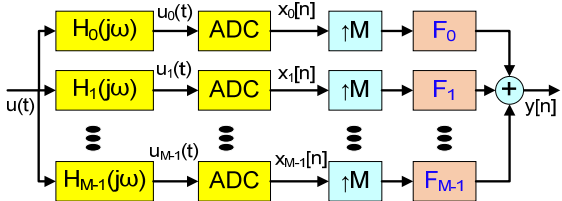

Fig. 3. The block diagram of correction

Comparing to Fig. 2, we can get the DFN(y[n]) as (5).

$$Y(e^{j\omega}) = \frac{1}{MT_S} \sum_{k}^{\infty} (U(j(\frac{\omega}{T_S} - \frac{2\pi k}{MT_S})) \times \sum_{m=0}^{M-1} (F_m(e^{j\omega}) H_m(j(\frac{\omega}{T_S} - \frac{2\pi k}{MT_S})))) \quad (5)$$

Considering the periodicity and symmetry of the spectrum of discrete digital signal, we usually take care of the spectrum in the base-band (i.e. ω ϵ [0, π)). So in (5), we only need to calculate the sum with finite number of the value of k. And k is determined by the limiting condition expressed as (6).

$$\left| \omega - \frac{2\pi k}{M} \right| < \pi \quad (6)$$

Although the value of k is related to the input frequency, the value of k which we need consider in (5) is finite and the number is only M. So the perfect reconstruct condition is described as (7) from which we can calculate the coefficient of the correction filters.

$$\Gamma_k = \sum_{m=0}^{M-1} F_m(e^{j\omega}) H_m(j(\frac{\omega}{T_S} - \frac{2\pi k}{MT_S}))$$

$$= \begin{cases} Me^{j\omega d}, & k = 0 \\ 0, & k = others \ M-1 \ integer \end{cases} \quad (7)$$

For an instance, in a four-channel TIADC system, the spectrum is shown as Fig. 4, and the value of k is also shown in it for the certain frequency ω.

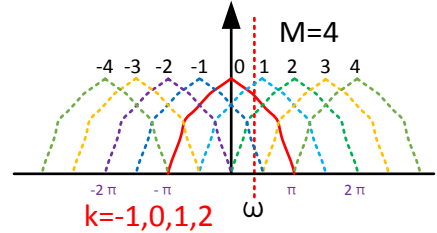

Fig. 4. The spectrum of TIADC digital output

#### 2) Wide band correction

In the narrow band normal sampling application, we can consider the mismatch constant and then we can calculate the responses of M filters and then get both the analytic expression of the filters and the detailed filter coefficients through IDFT conversion[11]. However, in the broad band application, we can't get the analytic solution. For this situation, we need another way to solve it.

Referring to Ref. 23, we can use the N points (N should be chosen large enough) pulse response to replace the filter response $F_m(e^{j\omega})$. In other words, we just need to solve $F_m(e^{j\omega})$ (ω = 2πn/N, n = 0, 1, 2…N-1), then we can get the filter responses within minimum mean square error. Here we can use this method to solve the perfect reconstruct condition.

Although the responses ($H_m(j\Omega)$) of input circuits are complex and we can't get definite analytic expressions, we can obtain the mismatches by interpolation through calibration.

First, we can measure the mismatches through injecting a sine wave signal to the TIADC system. Then the measurement process is repeated with a series of frequencies. After these, we can get the response of input circuits at any frequency by interpolation calculation.

After we get the input responses, we can calculate the filter response based on numerical solution by solving the M equations in (7) in a series of frequencies, 2πn/N (n = 0, 1, 2…N-1). N is the samples of input frequencies that we selected to calculate filter response.

Although we can't get the absolute solution of filter responses by numerical calculation, the accuracy is enough for common high-speed ADC resolution. Then we can get the filter coefficients under the least mean square error approximation through IDFT. An appropriate window is necessary to get a fewer order filter for the implementation in FPGA. The resolution is also applicable to the situation with constant mismatches and the solution is similar to the perfect analytic solutions.

### 3) Correction in Under-Sampling Situation

As under-sampling method is useful for reducing sampling rate, the correction in under-sampling for TIADC system is also important. When considering the under-sampling situation, we can still write the perfect reconstruct condition similar to (7), just with a few differences on the value of k.

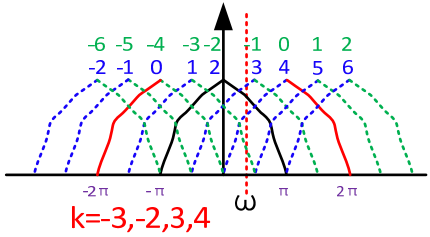

Fig. 5. The spectrum of TIADC digital output in under sampling

Considering the periodicity and symmetry, we just discuss it in the second Nyquist band in details. In this situation, the digital frequency range is $\omega \in [0, \pi)$, and the limited condition is presented as (8).

$$\pi < \left| \omega - \frac{2\pi k}{M} \right| < 2\pi \qquad (8)$$

In a four-channel TIADC system, the spectrum in under-sampling is shown as Fig. 5. And the value of k is also shown in it for a certain frequency $\omega$.

No matter the mismatches are constant or variable, it is difficult to get the analytic solutions. But the numerical solution method for the wide band correction is still useable. The processing is similar: firstly, calibrate the mismatches in the second Nyquist band, secondly solve the perfect reconstruct condition and thirdly design the filter banks.

### 4) Simulation Results

With the analysis above, we can simulate the correction process in Matlab. Fig. 6 and Fig. 7 are the typical spectrum before correction and after correction in normal sampling and under-sampling. We can obviously find that the mismatch errors are greatly suppressed by the correction process. Fig. 8 shows the simulation of the situation for two sine wave input signals. The simulation results show that the correction is working well for improving the performance of TIADC system within a wide range of frequency.

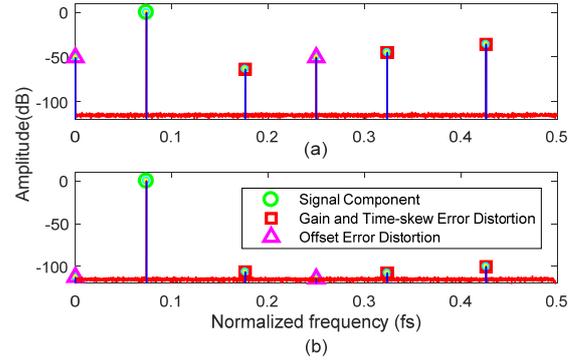

Fig. 6. Typical spectrum of simulation in normal sampling. (a) Before calibration; (b) After calibration

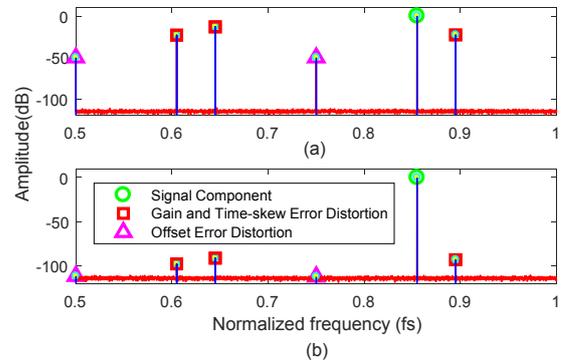

Fig. 7. Typical spectrum of simulation in under-sampling. (a) Before calibration; (b) After calibration

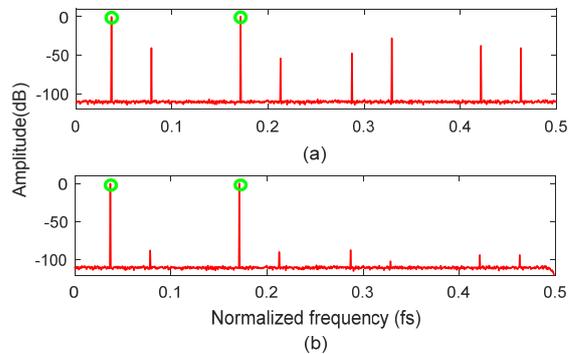

Fig. 8. Typical spectrum of simulation for two sine wave input signals in normal sampling. (a) Before calibration; (b) After calibration

After the verification by simulation, we need to evaluate the performance of the correction with the test data of TIADC system using software.

### III. PERFORMANCE TEST RESULT

After the analyzing and simulation, we have conducted a series of tests. To evaluate the performance of our correction algorithm, we use two different TIADC systems which had been designed: a 1.6-Gsps 14-bit and a 10Gsps 8-bit TIADC system. In our previous work, we conducted mismatch correction for a narrow band signal, which means that we had to change the coefficients of the filters when switching input signal frequency.[24] Apparently, there existed much limitation for real application. In this paper, we employed our new broad band correction algorithm and verified the performance through tests.

In the tests, we used a high performance RF signal source R&S SMA 100A to generate input test sinusoidal signals, which were further processed by external Band Pass Filters (BPFs) before fed to the TIADC system under test. We obtained the dynamic performance of these two systems based on the IEEE Std.1241-2010[25]. We implemented our correction algorithms in the MATLAB platform, with only one set of filter coefficients for input signal frequency in a wide band. Then we compared the system performance before and after correction.

## A. Results of the 1.6-Gsps 14-bit TIADC System

In the first TIADC system, four 14-bit ADCs work in parallel to obtain a system sampling rate up to 1.6 Gsps.

Fig. 9 shows the typical spectrum before and after correction with a 200 MHz input sine wave signal. As we observe, the distortions caused by mismatch errors are greatly suppressed.

To confirm the broad band correction effect, we employed a straight forward test method, and used two sinusoidal input signals to see whether the mismatch errors can be corrected. Fig. 10 shows the test results, which agrees well with the simulation results in Fig. 8.

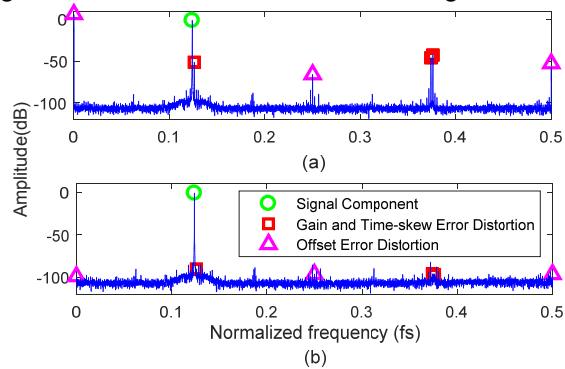

Fig. 9. Typical spectrum for single sine wave input signal in 1.6-Gsps 14-bit TIADC system. (a) Before calibration; (b) After calibration

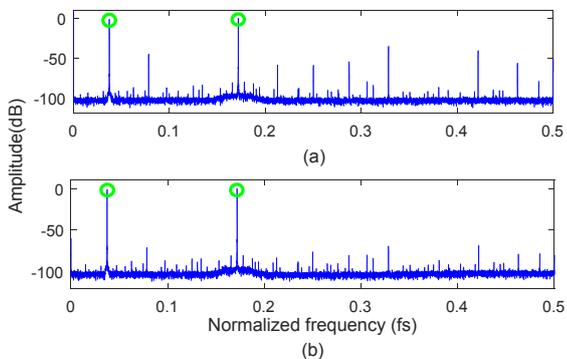

Fig. 10. Typical spectrum for two sine wave input signals in 1.6-Gsps 14-bit TIADC system. (a) Before calibration; (b) After calibration

To obtain the systematic performance of our correction method, we tuned the input signal frequency, and conducted a series of tests. As shown in Fig. 11, after correction, the ENOB performance is greatly enhanced to be better than 9.5 bits, which is close to the single ADC (ADS5474 from Texas Instruments Corporation) performance according to its datasheet. The results indicate that with the wideband correction algorithm proposed in this paper, good effect can be achieved over a wide input frequency range. We also plotted the performance of the narrowband perfect reconstruction correction algorithm[11] for comparison, as the black curve in Fig. 11. As shown in Fig. 11, the narrow band correction exhibits good effect at the frequency point of 60 MHz, but the performance deteriorates at other frequencies, especially with the increase of input signal frequency, and this is because the mismatch errors change significantly in high frequency range.

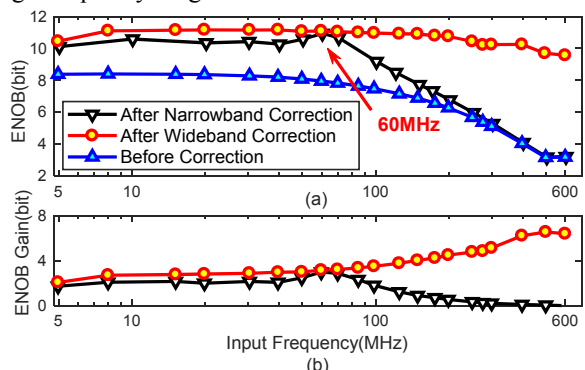

Fig. 11. ENOB test results of 1.6-Gsps 14-bit TIADC system in the frequency range from 5MHz to 600 MHz. (a) ENOB test results; (b) ENOB gain using different method.

After the test in the first Nyquist zone, we tested the TIADC performance in under-sampling situation. Due to the limitation of the ADC bandwidth (~600 MHz) in this TIADC system, we had to switch down the sampling clock frequency to 533.3 MHz. Fig. 12 shows the results of typical spectrum before and after correction, and it can be observed that the mismatch errors are reduced obviously. The Fig. 13 also indicates that the ENOB is significantly improved by the correction algorithm, and better than 9.5 bits in the second Nyquist zone (up to 500 MHz), which is almost the same with performance in the ADC datasheet.

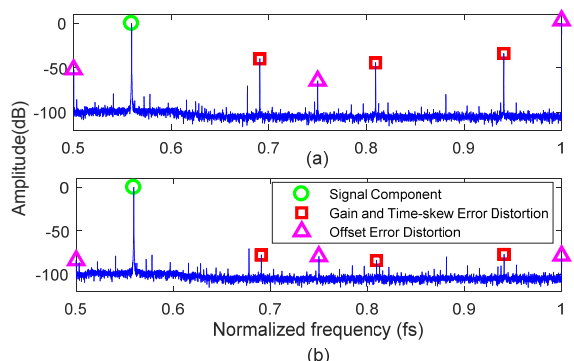

Fig. 12. Typical spectrum of 1.6-Gsps 14-bit TIADC system for single sine wave input signal in under-sampling situation. (a) Before calibration; (b) After calibration

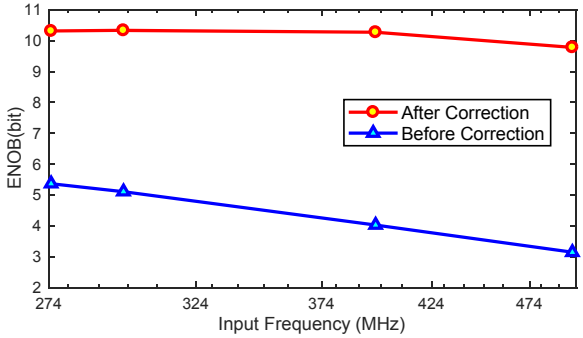

Fig. 13. ENOB test results of 1.6-Gsps 14-bit TIADC in under-sampling situation

## B. Results of the 10-Gsps 8-bit TIADC System

The 10-Gsps TIADC system consists of two 5-Gsps 8-bit EV8AQ160 Quad ADCs, each of which contains four 1.25-Gsps Sub-ADC cores inside.[26]

Fig. 14, Fig. 15, and Fig. 16 show the typical frequency spectrum with an input signal frequency of 200 MHz, frequency spectrum of the signal with two input signal frequencies, and the ENOB results before and after correction. We can observe that the system performance can be greatly improved by the correction algorithm, and the ENOB is better than 6.5 bits up to 1 GHz (7 bits from 20 MHz to 600 MHz), which is almost the same with the typical ADC performance according to its datasheet. Similarly, the ENOB deteriorates dramatically in high frequency range using the narrow frequency band correction method with which the variation of mismatches conducts poor correction effect in high frequency band.

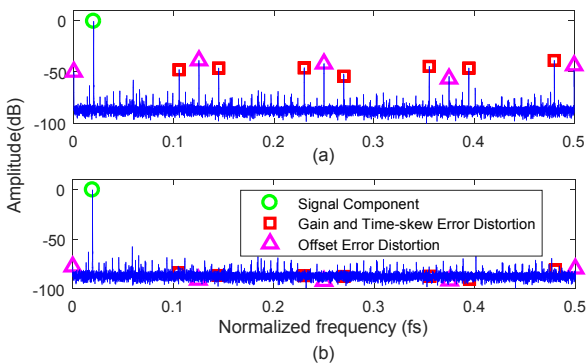

Fig. 14. Typical spectrum for single sine wave input signal in 10-Gsps 8-bit TIADC system. (a) Before calibration; (b) After calibration

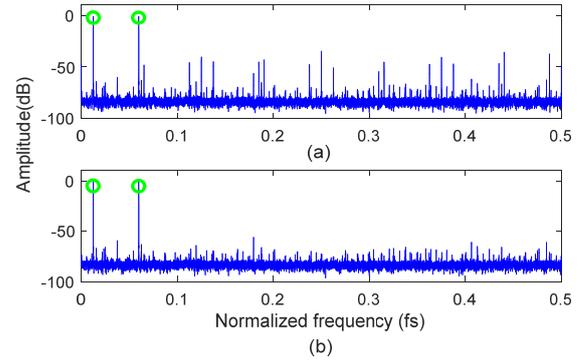

Fig. 15. Typical spectrum for two sine wave input signals in 10-Gsps 8-bit TIADC system. (a) Before calibration; (b) After calibration

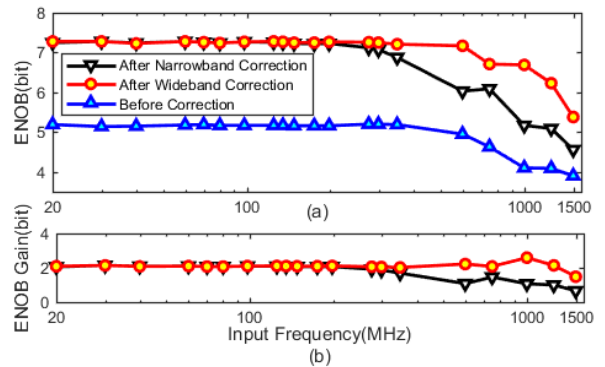

Fig. 16. ENOB test results of 10-Gsps 8-bit TIADC system in the frequency range from 20MHz to 1600 MHz. (a) ENOB test results; (b) ENOB gain using narrowband and wideband correction.

We tuned down the sampling speed of this TIADC system to 1428.6 Msps, in order to verify the correction performance in the under-sampling situation. As shown in Fig. 17 and Fig. 18, our correction algorithm achieves good effects in the second Nyquist zone, and ENOB is enhanced by 2 bits after correction, which is also close to single ADC performance. Of course, according to the architecture of our correction algorithm, the correction is not only limited to the first and second Nyquist zones, but also can be applied in other frequency bands.

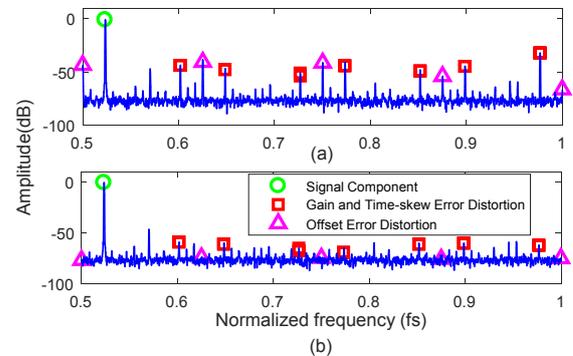

Fig. 17. Typical spectrum of 10-Gsps 8-bit TIADC for single sine wave input signal in under-sampling situation. (a) Before calibration; (b) After calibration

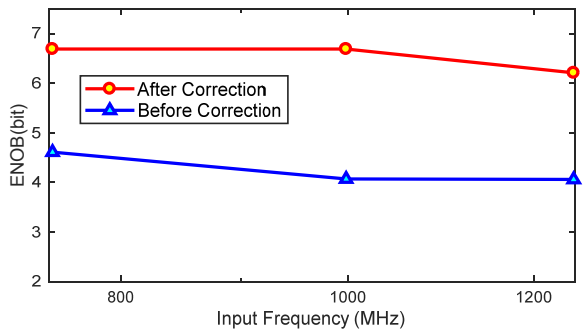

Fig. 18. ENOB test results of 10-Gsps 8-bit TIADC in under-sampling situation.

## IV. CONCLUSION

We present a new fully digital mismatch correction method for TIADC systems over a wide-band frequency range. Studies were made to cover the base-band and under sampling situation. The correction algorithm performance was further verified through tests on two TIADC systems: a 1.6-Gsps 14-bit and a 10-Gsps 8-bit system. Test results indicate that this correction algorithm can obviously improve the performance of TIADC system over a broad frequency band both in the first and second Nyquist zones.

## ACKNOWLEDGMENTS

This work was supported in part by the National Natural Science Foundation of China under Grant 11675173, in part by the Knowledge Innovation Program of the Chinese Academy of Sciences under Grant KJCX2-YW-N27, and in part by the CAS Center for Excellence in Particle Physics (CCEPP).

[1] X. Hu, L. Zhao, W. Zheng, S. Liu, and Q. An, "Data acquisition system based on Time-Interleaved Analog-to-Digital Conversion for Time-of-Flight Mass Spectrometer," in Real Time Conference (RT), 2012 18th IEEE-NPSS, 2012, pp. 1-7.

[2] Liu, Jin, and Vinod Mukundagiri. "Study of ADC resolution and bandwidth requirement tradeoffs for high-speed data communications." Circuits and Systems (MWSCAS), 2013 IEEE 56th International Midwest Symposium on. IEEE, 2013.

[3] Zhao, Xiaodong, et al. "GPS-Disciplined Analog-to-Digital Converter for Phasor Measurement Applications." IEEE Transactions on Instrumentation and Measurement (2017).

[4] Joly, Baptiste, Gérard Montarou, and Pierre-Etienne Vert. "Sampling Rate and ADC Resolution Requirements in Digital Front-End Electronics for TOF PET." IEEE Transactions on Nuclear Science (2017). .

[5] Duan Y, Alon E, "A 12.8 Gs/s time-interleaved adc with 25 GHz effective resolution bandwidth and 4.6 enob," IEEE Journal of Solid-State Circuits, 2014, 49(8), pp. 1725-1738.

[6] Seo M, Rodwell M J W, Madhow U, "A low computation adaptive blind mismatch correction for time-interleaved ADCs," Circuits and Systems, 2006. MWSCAS'06. 49th IEEE International Midwest Symposium on. IEEE, 2006, 1, pp. 292-296.

[7] Wang, Xiao, Fule Li, and Zhihua Wang. "A novel autocorrelation-based timing mismatch C alibration strategy in Time-Interleaved ADCs." Circuits and Systems (ISCAS), 2016 IEEE International Symposium on. IEEE, 2016.

[8] Selva J. "Functionally weighted Lagrange interpolation of band-limited signals from nonuniform samples," IEEE Transactions on Signal Processing, 2009, 57(1), pp. 168-181.

[9] Johansson H, Lowenborg P. "Reconstruction of nonuniformly sampled bandlimited signals by means of digital fractional delay filters," IEEE Transactions on signal processing, 2002, 50(11), pp. 2757-2767.

[10] Dilmaghani, Mehdi Sefidgar, and Davud Asemani. "Correction of timing skew error in TIADC's using second generation wavelets." Electrical Engineering (ICEE), 2017 Iranian Conference on. IEEE, 2017.

[11] Lee Y S, An Q. "Calibration of time-skew error in a M-channel time-interleaved analog-to-digital converter," World Academy of Science, Engineering and Technology, 2005, 2, pp. 208.

[12] Saleem S, Vogel C. "Adaptive compensation of frequency response mismatches in high-resolution time-interleaved ADCs using a low-resolution ADC and a time-varying filter," Circuits and Systems (ISCAS), Proceedings of 2010 IEEE International Symposium on. IEEE, 2010, pp. 561-564.

[13] Jin H, Lee E K F. "A digital-background calibration technique for minimizing timing-error effects in time-interleaved ADCs," IEEE Transactions on Circuits and Systems II: Analog and Digital Signal Processing, 2000, 47(7), pp. 603-613.

[14] Saleem S, Vogel C. "Adaptive blind background calibration of polynomial-represented frequency response mismatches in a two-channel time-interleaved ADC," IEEE Transactions on Circuits and Systems I: Regular Papers, 2011, 58(6), pp. 1300-1310.

[15] Liu, Husheng, and Hui Xu. "An Adaptive Blind Frequency-Response Mismatches Calibration Method for Four-Channel TIADCs Based on Channel Swapping." IEEE Transactions on Circuits and Systems II: Express Briefs 64.6 (2017): 625-629.

[16] Matsuno J, Yamaji T, Furuta M, et al., "All-digital background calibration technique for time-interleaved ADC using pseudo aliasing signal," IEEE Transactions on Circuits and Systems I: Regular Papers, 2013, 60(5), pp. 1113-1121.

[17] Jamal S M, Fu D, Singh M P, et al., "Calibration of sample-time error in a two-channel time-interleaved analog-to-digital converter," IEEE Transactions on Circuits and Systems I: Regular Papers, 2004, 51(1), pp. 130-139.

[18] Jamal S M, Fu D, Chang N C J, et al. "A 10-b 120-Msample/s time-interleaved analog-to-digital converter with digital background calibration," IEEE Journal of Solid-State Circuits, 2002, 37(12), pp. 1618-1627.

[19] Mendel S, Vogel C. "A compensation method for magnitude response mismatches in two-channel time-interleaved analog-to-digital converters," Electronics, Circuits and Systems, 2006. ICECS'06. 13th IEEE International Conference on. IEEE, 2006, pp. 712-715.

[20] Shahmansoori A. "Adaptive blind calibration of timing offsets in a two-channel time-interleaved analog-to-digital converter through Lagrange interpolation," Signal, Image and Video Processing, 2015, 9(5), pp. 1047-1054.

[21] Le Duc H, Nguyen D M, Jabbour C, et al. "Hardware implementation of all digital calibration for undersampling TIADCs," Circuits and Systems (ISCAS), 2015 IEEE International Symposium on. IEEE, 2015, pp. 2181-2184.

[22] Le Duc, Han, et al. "Fully Digital Feedforward Background Calibration of Clock Skews for Sub-Sampling TIADCs Using the Polyphase Decomposition." IEEE Transactions on Circuits and Systems I: Regular Papers 64.6 (2017): 1515-1528.

[23] Velazquez S R, Nguyen T Q, Broadstone S R. "Design of hybrid filter banks for analog/digital conversion," IEEE transactions on signal processing, 1998, 46(4), pp. 956-967.

[24] Zhao L, Hu X, Feng C, et al. "A 1.6-gsps high-resolution waveform digitizer based on a time-interleaved technique," IEEE Transactions on Nuclear Science, 2013, 60(3), pp. 2180-2187.

[25] IEEE Standard for Terminology and Test Methods for Analog-to-Digital Converters, IEEE Standard 1241-2010, Jan. 2011.

[26] Shaochun Tang, "Research of Ultra-High-Speed High-Resolution Waveform Digitization Based on Time-Interleaved Technique," Ph.D. dissertation, Univ. Science and Technology of China, Hefei, 2012, pp. 77-85